\documentclass{article}
\usepackage[T1]{fontenc} 
\usepackage[utf8]{inputenc} 
\usepackage{ismir,amsmath,cite,url}
\usepackage{graphicx}
\usepackage{color}
\usepackage{booktabs}

\usepackage{lineno}

\title{Evaluating the Evaluation of Music Generation Models with Generative Psychometrics}

\multauthor
{ João Carlos Santana Silveira $^1$ \hspace{1cm} Caio Von Rondow Morais$^1$ \hspace{1cm} Frederico Pedrosa$^2$}
{\bfseries{Lucas N. Ferreira$^1$}\\
$^1$Departamento de Ciência da Computação, UFMG -- Belo Horizonte, MG, Brazil \\
$^2$Escola de Música, UFMG -- Belo Horizonte, MG, Brazil \\
{\tt\small \{joaocarlossantana, caiorondow, lferreira\}@dcc.ufmg.br, fredericopedrosa@musica.ufmg.br}
}
\def\authorname{J. Silveira, C. Morais, F. Pedrosa, L. Ferreira}

\begin{document}

\maketitle
\begin{abstract}
Subjective user studies are one of the major methods for evaluating music generation models. However, subjective evaluation protocols vary widely and are rarely specified in enough detail to be reproduced, hindering comparison across studies. In this paper, we map subjective evaluation methods in ISMIR music generation papers from 2020–2025 and apply generative psychometrics to assess the stability of their items. We extract 284 items from the papers’ methodological descriptions, using an LLM and manual validation to standardize them. We represent these items as contextual embeddings and apply Unique Variable Analysis to remove redundant, locally dependent items and Bootstrap Exploratory Graph Analysis to estimate their latent dimensions. Results show the network converges onto a taxonomy of six stable quality dimensions: Rhythm, Melody, Audio, Composition, Structure, and Lyrics, pointing to a unified methodology for conducting user studies to evaluate music generation models.\end{abstract}
\section{Introduction}\label{sec:introduction}

Generative models of music have advanced considerably over the last decade, from early unconditioned recurrent neural networks that produced symbolic music with little long-term coherence \cite{Oore2017}, to modern multimodal transformers that generate complete audio pieces with strong long-term structure, conditioned on text \cite{copet2023simple} or video \cite{tian2024vidmuse}. This progress has been assessed through two complementary evaluation methodologies: objective metrics, which capture structural qualities of the generated pieces, and listening tests, which capture the perceived quality of the music as judged by human listeners. Objective metrics have evolved alongside the models but remain relatively consistent across studies; subjective protocols have not kept the same consistency and are often left underspecified and therefore hard to reproduce. This wide variation in subjective evaluation protocols hinders comparison across studies, slowing progress in the field.

In this paper, our goal is to identify patterns of evaluation dimensions (e.g., melody, harmony, rhythm) in subjective listening tests, as well as representative items (specific questions) within each dimension, in order to propose a unified protocol. To this end, we map subjective evaluation methods from the music generation papers published at the International Society for Music Information Retrieval (ISMIR) Conference between 2020 and 2025. We start by retrieving all ISMIR papers from this period and filtering the corpus down to papers strictly related to music generation, which is done by a Large Language Model (LLM) and validated manually. We then manually extract the items used in their subjective evaluations. Since these items are often not stated verbatim in the papers, we use an LLM to standardize them following guidelines from the psychometrics literature \cite{itc2001guidelines}. This standardization process is manually validated in an item cleaning step, where we compare the original item descriptions from the papers with the items generated by the LLM to correct inconsistencies. During this step, we remove generated items that were hallucinated by the LLM due to the lack of detail in the original papers. In the end, this entire process yields a total of 284 standardized items.

The resulting items are analyzed with a generative psychometrics method called AI-GENIE (Automatic Item Generation and Validation with Network-Integrated Evaluation), which first extracts contextual embeddings of the items to project them into the same vector space and then applies Exploratory Graph Analysis (EGA) to estimate a sparse network in which nodes represent items and edges represent the strength of their pairwise semantic similarity. Following the generative psychometrics literature \cite{liu2020domain, oneill2020meta}, we embed our items with the OpenAI \texttt{text-embedding-3-small} model.

Before communities are formed, Unique Variable Analysis (UVA) is applied to this network to detect and remove redundant items, which narrows the 284 items down to 57 unique items. The purified network is then partitioned into communities using the Walktrap algorithm, which performs short random walks across the graph to locate groups of densely interconnected items. Each resulting community is interpreted as an empirically derived latent dimension, so that the item pool converges onto a small set of underlying constructs based on how strongly their embeddings interconnect.

In our results, the 57 unique items were grouped into six stable communities, which were analyzed by one of the authors, an assistant professor of music, to assign semantic labels to them. This analysis found that the communities grouped the items according to the following quality dimensions: Rhythm, Melody, Audio, Composition, Structure, and Lyrics. Moreover, a centrality analysis automatically identified the most representative items within each dimension; together, these results point to a unified methodology for conducting user studies to evaluate music generation models. The data, source code, and LLM prompts to reproduce our results are available on GitHub\footnote{\url{https://github.com/jcssilveira/music-gen-eval}}.

\section{Related Work}\label{sec:related}

\subsection{Music Generation Models}

Early efforts to generate music with neural networks focused on the symbolic domain, where music is represented as discrete events (e.g., notes, pitches, durations) rather than raw audio waveforms. For example, Oore et al. \cite{Oore2017} used an LSTM recurrent network to generate expressive piano performances, capturing note-level dynamics and timing but lacking high-level structure over longer passages. Huang et al. \cite{huang2018music} addressed this with the Music Transformer, using relative self-attention to preserve thematic and structural coherence across much longer sequences. This shift from recurrent to attention-based architectures underlies the large-scale music generation systems that followed.

As audio-generation techniques improved, the field moved from symbolic representations to modeling audio directly, capturing timbre and vocal content alongside musical structure. For example, Dhariwal et al. \cite{dhariwal2020jukebox} introduced Jukebox, combining a hierarchical VQ-VAE with autoregressive transformers to generate multi-minute songs with lyrics-conditioned singing. Subsequent work added multimodal conditioning: MusicLM \cite{Agostinelli2023} and MusicGen \cite{copet2023simple} generate audio from free-form text descriptions, while Video2Music \cite{kang2024video2music} and VidMuse \cite{tian2024vidmuse} generate music conditioned on video.

\subsection{Evaluation}

Objective evaluation relies on automatically computable metrics that approximate perceptual quality without human raters. For symbolic music, feature-based metrics compare score-level statistics between generated and reference corpora: examples include the polyphony and pitch-in-scale rates introduced with MuseGAN \cite{dong2018musegan}, scale consistency from C-RNN-GAN \cite{mogren2016c}, and pitch-class entropy and groove consistency from the Jazz Transformer \cite{wu2020jazz}, commonly computed with toolkits such as MusPy \cite{dong2020muspy}. For audio-domain models, the Fréchet Audio Distance (FAD) \cite{kilgour2018fr} compares distributions of embeddings extracted from generated and real recordings and has become a standard measure of audio fidelity and diversity. Text-conditioned models are additionally assessed with the CLAP score \cite{elizalde2023clap}, which uses a contrastive audio-text embedding space to quantify how well generated audio matches its conditioning description.

Despite this growth in automatic metrics, listening tests remain the primary means of validating perceptual quality, since they capture attributes such as musicality and emotional impact that objective metrics cannot. Most studies adopt Mean Opinion Score (MOS) protocols, where participants rate stimuli on Likert scales along dimensions such as overall quality or naturalness, or the MUSHRA methodology \cite{11460460}, which compares multiple stimuli against a hidden reference and anchor. These protocols, however, are rarely applied consistently: the number of participants, rating scales, and evaluation dimensions vary considerably across papers, motivating the systematic mapping undertaken in this work.

\subsection{Generative Psychometrics}

Generative psychometrics is an emerging paradigm that leverages LLMs for item generation and structural validation of measurement instruments \cite{russell2026generative, varrasi2026expanding}, building on evidence that LLM embedding spaces align with human response covariation \cite{kambeitz2025empirical}. Its methodological core is AI-GENIE, which draws on distributional semantics to approximate human covariation patterns from the semantic proximity between item descriptions \cite{varrasi2026expanding}. By isolating items' semantic content, AI-GENIE bypasses situational noise inherent to human data (e.g., fatigue, misinterpretation, social desirability bias), recovering an instrument's latent structure prior to any empirical data collection \cite{garrido2025estimating}.

Structural validation is achieved through a network psychometrics pipeline: UVA iteratively prunes semantically redundant items until each remaining node contributes unique variance, and Bootstrap Exploratory Graph Analysis (bootEGA) resamples the data to detect and remove unstable items, converging once the surviving items show high topological stability. In the musical domain, Ribeiro et al. \cite{ribeiro2026arquitetura} applied this approach to literature-derived items and predicted a three-dimensional structure for Musical Development (Musical Production, Musical Perception, and Music and Movement), consistent with the structure found empirically by Buren et al. \cite{buren2025screening} via exploratory and confirmatory factor analyses.

\section{Generative Psychometrics for Music Generation Items}\label{sec:Method}

\begin{figure}[t]
    \centering
    \includegraphics[width=3.45cm]{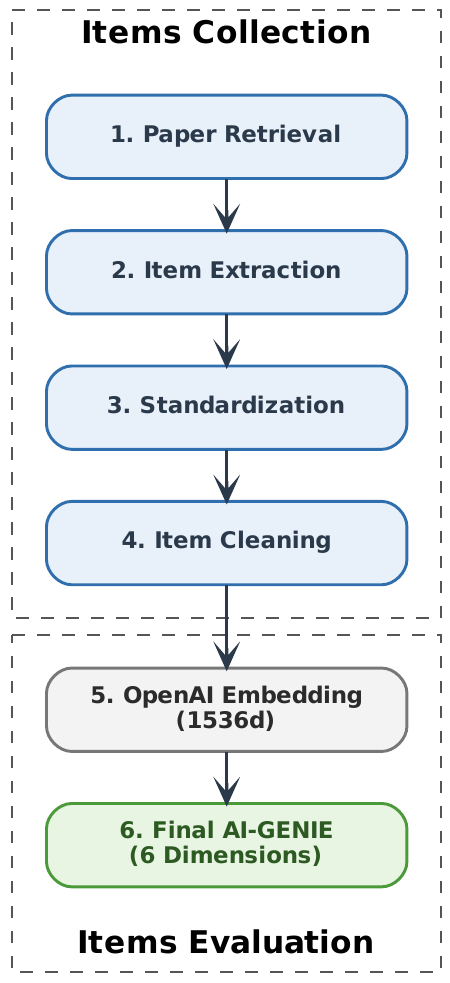}
    \caption{Six-step methodological pipeline of the study.}
    \label{fig:method_pipeline}
\end{figure}

The methodology of this study follows a six-step pipeline, as illustrated in Figure~\ref{fig:method_pipeline}: (1) retrieval of music generation papers; (2) extraction of the raw subjective evaluation questions; (3) LLM standardization of the items; (4) item cleaning to ensure consistency between standardized and raw items; (5) embedding extraction; and (6) the final AI-GENIE execution to derive the network with stable dimensions. These steps can be organized into two parts: Item Collection, from steps (1) to (4), and Item Evaluation, from steps (5) to (6).

\subsection{Item Collection}

The collected papers comprise the publications in the ISMIR conference proceedings between 2020 and 2025. As an initial base, we utilized the ISMIR25Meta dataset \cite{gomez2025beyond}, removing the papers published before 2020, since our focus is to extract item patterns from relatively recent evaluation protocols. Then, the papers from the 2025 edition were manually added to the final set. This step resulted in 648 papers, all with their abstracts, whose annual distribution consists of: 115 (2020), 104 (2021), 112 (2022), 103 (2023), 123 (2024), and 91 (2025).

The next step consisted of identifying which of these 648 papers were strictly related to music generation. For that, we applied an LLM (Gemini) to classify, based on the abstracts, whether or not each paper addressed generation tasks. This process identified 150 papers (a reduction of 77\% of the original set). In parallel, we conducted a keyword-based analysis over the 648 papers as a way to validate the LLM. This method returned a subset of 152 papers, corroborating the LLM classification. The final subset of 150 music generation papers presents the following annual distribution: 24 (2020), 21 (2021), 31 (2022), 18 (2023), 36 (2024), and 20 (2025).

With that subset in hand, we performed a manual extraction of the subjective evaluation sections detailing how user studies were conducted. At this stage, we focused on mapping which questions were included in the questionnaires, as well as the investigated metrics. Additionally, participant demographic data was collected. During this process, we observed that the majority of authors do not explicitly describe the exact questions used in their evaluations; in fact, only 5 out of the 150 papers provided working links to the original questionnaires.

After extracting the raw text from the listening test subsections, we employed an LLM to standardize the raw items, following established guidelines for test use and item adaptation \cite{itc2001guidelines}. For example, given the following excerpt from a paper's evaluation section:
\begin{quote}
``5.2 Subjective metrics include creativity (C), naturalness (N), and musicality (M) [...] and add a fitness (F) metric to evaluate how well the generated parts fit in with the given melody.''
\end{quote}
the LLM produced the following standardized items:
\begin{itemize}
    \item \textbf{Original Item}: ``creativity (C)'' \\
    \textbf{Standardized Item}: ``rate the generation quality in terms of creativity'' \\
    \textbf{Response Scale}: 5-point scale \\
    \textbf{Sample Size}: 36 participants
    \item \textbf{Original Item}: ``naturalness (N)'' \\
    \textbf{Standardized Item}: ``rate the generation quality in terms of naturalness'' \\
    \textbf{Response Scale}: 5-point scale \\
    \textbf{Sample Size}: 36 participants
    \item \textbf{Original Item}: ``musicality (M)'' \\
    \textbf{Standardized Item}: ``rate the generation quality in terms of musicality'' \\
    \textbf{Response Scale}: 5-point scale \\
    \textbf{Sample Size}: 36 participants
\end{itemize}

To mitigate potential hallucinations and ensure data reliability, we conducted a manual item cleaning step after the LLM standardization. We compared each raw item against its standardized counterpart and manually adjusted any inconsistencies, removing items that were hallucinated by the LLM due to the lack of clarity in the raw item descriptions. In the end, this process yielded a total of 284 standardized items.

\subsection{Item Evaluation}

For each of the 284 standardized items, we generated a contextual embedding using the OpenAI \texttt{text-embedding-3-small} model, which produces vectors of 1536 dimensions. The embeddings were then submitted to AI-GENIE to estimate the latent dimensionality of the item pool. UVA pruned the 284 items down to 57 unique items, which were then partitioned into communities via EGA with the Walktrap algorithm, and validated for stability with bootEGA. This process converged on six stable communities, each interpreted as an empirically derived latent dimension of subjective music generation evaluation. To ensure interpretability and domain relevance, each community was qualitatively reviewed and named by one of the authors---a professional musician and professor holding a Ph.D. in Music---who synthesized the core semantic content of the items comprising each dimension.

Following the establishment of this final structure, a centrality analysis was conducted to identify the most representative items within each dimension. In network psychometrics, an item's representativeness is primarily quantified through its network loading (NL), i.e., its standardized node strength within its own community, which indicates how strongly the item reflects its underlying construct.

\section{Results}\label{sec:results}

Through the iterative application of UVA, all local dependencies were eliminated, ensuring that every retained item contributes unique variance to the model, without semantic redundancy. Furthermore, the structural robustness of the final structure was confirmed through bootEGA. Following the resampling iterations, the final network achieved strong structural integrity, in which all items demonstrated stability between 84\% and 100\% across the bootstrap replications. This high stability indicates that the items consistently cluster within their designated dimensions, free from topological wandering. Ultimately, the AI-GENIE methodology successfully distilled the initial pool of heterogeneous questions into a structurally stable taxonomy for evaluating music generation models.

\begin{figure}[t]
    \centering
    \includegraphics[width=\columnwidth]{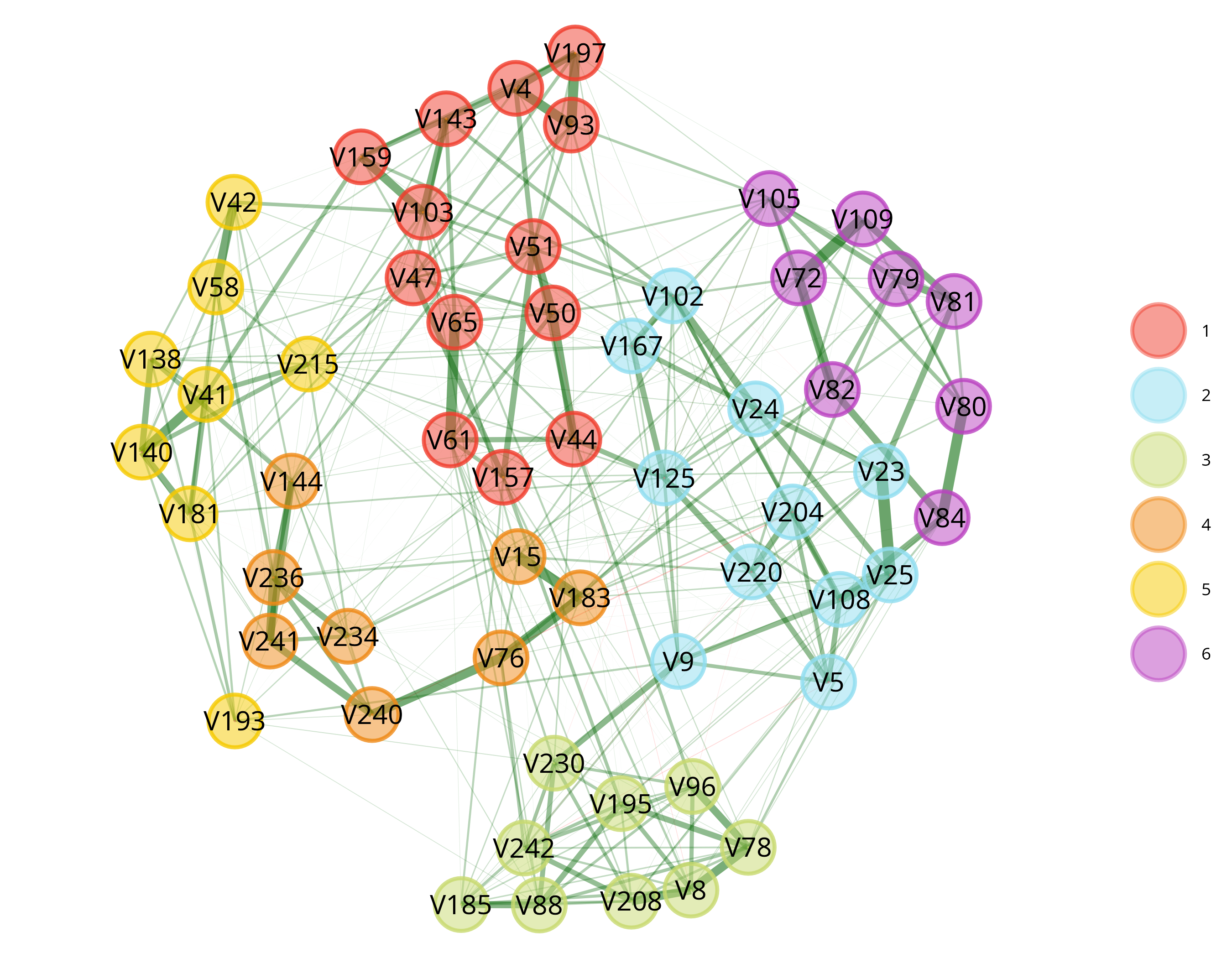}
    \caption{Final six-dimensional network structure estimated via EGA after item purification. Node colors indicate the six communities, node labels (V$x$) are item IDs, and edge thickness represents the strength of the association between items.}
    \label{fig:ega_final}
\end{figure}

Figure~\ref{fig:ega_final} illustrates the resulting network derived from the embeddings. In the following, the values in parentheses after each item ID indicate its network loading (NL). For Community 1, items V51 (NL = 0.59) and V143 (NL = 0.46) were the strongest anchors, prompting listeners to evaluate ``rhythmic stability (whether the drummer follows a steady tempo) of the generated drum'' and ``the variety of rhythm,'' respectively. Therefore, this community was labeled \textit{Rhythm}.

Community 2 was best represented by V108 (0.55) and V25 (0.48), which focus on assessing ``what is the overall quality of the melody?'' and ``the overall music quality of the melody reduction.'' Thus, this community was labeled \textit{Melody}. Meanwhile, the third dimension (Community 3) was strongly captured by V78 (0.53) and V8 (0.52), with criteria requesting participants to provide ``a subjective assessment of the audio quality'' and to rate overall ``audio quality.'' Therefore, Community 3 was named \textit{Audio}.

\begin{table}[b]
\centering
\small
\begin{tabular}{lp{4.25cm}l}
\toprule
\textbf{Dimension} & \textbf{Most Representative Item} & \textbf{NL} \\
\midrule
Rhythm & ``Rhythmic stability (whether the drummer follows a steady tempo) of the generated drum.'' & 0.59 \\
Melody & ``What is the overall quality of the melody?'' & 0.55 \\
Audio & ``A subjective assessment of the audio quality.'' & 0.53 \\
Composition & ``I like this composition.'' & 0.52 \\
Structure & ``Long-term structure.'' & 0.61 \\
Lyrics & ``Are the impressions of the input text and the generated lyrics similar?'' & 0.50 \\
\bottomrule
\end{tabular}
\caption{The six evaluation dimensions and their most representative item, with its network loading (NL).}
\label{tab:dimensions}
\end{table}

In the fourth dimension (Community 4), items V236 (0.52) and V241 (0.49) exhibited the highest loadings, primarily evaluating compositional reception and characteristics such as ``I like this composition'' and whether ``this composition is repetitive.'' Hence, this community was labeled \textit{Composition}. We note, however, that its strongest item captures the listener's overall preference rather than a specific compositional attribute, suggesting that this dimension combines subjective liking with judgments about the composition itself. Community 5 displayed the strongest individual loading in the entire network, with item V140 (0.61), which evaluates ``long-term structure,'' followed by V41 (0.56), which gauges ``structureness.'' This community was labeled \textit{Structure}. Finally, Community 6 was anchored by V72 (0.50) and V82 (0.49), reflecting questions such as ``Are the impressions of the input text and the generated lyrics similar?'' and ``How singable are the generated lyrics?'' Thus, Community 6 was named \textit{Lyrics}.

Overall, these results show that there are six major dimensions of evaluation within the music generation papers published at ISMIR between 2020 and 2025. Notably, harmony did not emerge as an independent dimension, despite being a core musical attribute; harmony-related items were either removed as redundant by UVA or absorbed into broader communities such as \textit{Melody} and \textit{Composition}. These dimensions, as well as their most representative items, are summarized in Table~\ref{tab:dimensions}.


\section{Conclusion}
This paper mapped the heterogeneous landscape of subjective evaluation protocols in music generation research to identify a stable set evaluation dimensions. Applying generative psychometrics to 284 items extracted from 150 ISMIR papers (2020--2025), we found that this heterogeneity converges onto six stable latent dimensions: \textit{Rhythm}, \textit{Melody}, \textit{Audio}, \textit{Composition}, \textit{Structure}, and \textit{Lyrics}.

As future work, since this study is limited to a single venue and constrained by the scarcity of publicly available questionnaires, one could empirically validate these dimensions with new listening tests and extend the analysis to other venues. One could also compare the performance of different embedding models, as this paper focused on a single model.



\bibliography{ISMIRtemplate}

\end{document}